\documentclass{iopart}
\usepackage{iopams}  
\usepackage{graphicx}

\newcommand{\be}[0]{
	\begin{equation}
}

\newcommand{\ee}[0]{
	\end{equation}
}

\begin{document}

\title{Jump probabilities in the non-Markovian quantum jump method}

\author{Kari H\"{a}rk\"{o}nen}

\address{Turku Centre for Quantum Physics, Department of Physics and Astronomy, University of Turku, FI-20014 Turku, Finland}

\ead{kari.harkonen@utu.fi}

\begin{abstract}
The dynamics of a non-Markovian open quantum system described by a general time-local master equation is studied. The propagation of the density operator is constructed in terms of two processes: (i) deterministic evolution and (ii) evolution of a probability density functional in the projective Hilbert space. The analysis provides a derivation for the jump probabilities used in the recently developed non-Markovian quantum jump (NMQJ) method [Piilo~J \textsl{et al} 2008 \textsl{Phys.~Rev.~Lett.}~\textbf{100} 180402]. 
\end{abstract}

%Uncomment for PACS numbers title message
\pacs{03.65.Yz, 42.50.Lc}
% Keywords required only for MST, PB, PMB, PM, JOA, JOB? 
%\vspace{2pc}
%\noindent{\it Keywords}: Article preparation, IOP journals
% Uncomment for Submitted to journal title message
%\submitto{\JPA}
% Comment out if separate title page not required
% \maketitle

\section{Introduction}

The theory of open quantum systems \cite{Breuer2007} provides means to study in detail a selected part of the total Hilbert space. This is done by making a division into a system part and the rest of the space, which can be seen as an environment to the system of interest. The position of the artificial boundary between the two parts, also referred to as the Heisenberg cut~\cite{Strunz1999}, can be selected arbitrarily according to the problem in hand. The interaction of the system part with its environment appears as non-coherent dynamics, and the evolution of the reduced system has to be described by a master equation. 

Non-Markovian dynamics carries a trace from the past. In the open quantum systems, non-Markovianity emerges from the enforced reduction of state space. Non-Markovian dynamics has been encountered in many fields of physics, such as quantum optics \cite{Gardiner1999}, solid state physics \cite{Lai2006}, quantum chemistry \cite{Shao2004}, quantum information processing \cite{Aharonov2006}, in the biological context \cite{Thorwart2009,Rebentrost2009}, and even as a resource to manipulate the quantum-classical border \cite{Maniscalco2006}. Solving the non-Markovian dynamics, even numerically, is considerably more complicated than in the Markovian case. 

Two standard types of non-Markovian master equations are (i) the Nakazima-Zwanzig equation \cite{Nakajima1958,Zwanzig1960}, which is an integro-differential equation including a time convolution of the state history with a memory-kernel, and (ii) time-local expressions \cite{Shibata1977,Chaturvedi1979}. In a memory-kernel master equation, the past evolution is clearly present, whereas in a time-local expression there is no explicit dependence on the history. Consequently, it appears counterintuitive that the time-local master equations could produce non-Markovian dynamics at all. Time-convolutionless (TCL) projection operator techniques~\cite{Breuer2007,Shibata1977,Chaturvedi1979} provide the general mathematical machinery to form time-local master equations. 

Recently, non-Markovian quantum jump (NMQJ) method \cite{Piilo2008,Piilo2009} was introduced as a simulation algorithm for the dynamics of non-Markovian open quantum systems which are described by a general time-local master equation. It provides the widely used Monte Carlo wave function (MCWF) method \cite{Dalibard1992,Dum1992,Carmichael1993} with an extension to the non-Markovian regime. In contrast to other quantum-jump-based methods available for non-Markovian dynamics \cite{Garraway1997,Breuer1999,Breuer2004}, the NMQJ method is strictly confined to the original Hilbert space of the reduced system without any auxiliary degrees of freedom. 

The NMQJ method introduces the concept of a reverse quantum jump during the periods where decay rates, which are associated with the quantum-jump probabilities in the MCWF method, reach negative values. During such periods the system reabsorbs information and energy from its environment as decoherence and dissipation are reversed. Such new kind of a stochastic process occurs with a peculiar ensemble-dependent probability~\cite{Piilo2008,Piilo2009}. This paper provides a detailed line of reasoning to derive the jump probabilities used in the NMQJ method, and introduces ensemble-dependent jump operators which generate the reverse jumps.

The paper is organized as follows. In Sec.~\ref{sec:evolution} the evolution of the density operator in terms of a pure-state decomposition is constructed. Section~\ref{sec:connection} concentrates on analysing the jump-like processes and shows the connection to the quantum-jump probabilities used in the NMQJ method. The results are discussed in Sec.~\ref{sec:discussion}, and Sec.~\ref{sec:summary} summarizes and concludes the paper.

\section{\label{sec:evolution}Evolution}

The most general form of a time-local master equation for the reduced density operator $\rho(t)$ achievable by the TCL procedure is \cite{Breuer2009} 
\begin{eqnarray}
\fl \frac{ d \rho(t)}{ d t } = -\frac{ \rmi }{ \hbar } [ H(t), \rho(t) ] + \sum_{k} \Delta_k(t) \Big( C_k (t)\rho(t) C_k^\dagger (t) - \frac{1}{2} \big\{ C_k^\dagger (t) C_k (t), \rho (t) \big\} \Big),
\label{eq:masterEquation}
\end{eqnarray}
where the coherent dynamics is generated by the Hermitian Hamiltonian $H(t)$ (including possible Lamb and Stark shifts), dissipation and decoherence are induced by the Lindblad (jump) operators $C_k (t)$, and each decay channel $k$ is equipped with a decay rate $\Delta_k (t) \in \mathbb{ R }$. The above equation is a time-dependent generalization of the most general form of a Markovian master equation (where $\Delta_k>0$) generating completely positive dynamical maps \cite{GoriniA,LindbladB}. The master equation~\eref{eq:masterEquation} is local in time, i.e., it has no explicit dependence on the past evolution of the density operator. However, allowing the decay rates to enter negative values produces non-Markovian dynamics. 

For (time-dependent) Markovian systems, where $\Delta_k (t) > 0$ always, there exists a piecewise deterministic process for state vectors $| \psi \rangle$ such that the density operator given by a pure-state decomposition 
\be
\rho (t) = E\big[ | \psi \rangle \langle \psi | \big] = \int \rmd \psi P[\psi; t] \rho_\psi
\label{eq:density}
\ee
satisfies the master equation~\eref{eq:masterEquation} \cite{Breuer1995a,Breuer1995b}. Above, $\rmd \psi = \mathrm{D} \psi \mathrm{D} \psi^*$ is a singular volume element of the Hilbert space $\mathcal{H}$, and $P[\psi; t]$ is a time-dependent probability density functional on the projective Hilbert space $\mathcal{P} ( \mathcal{H} )$. The probability density $P[\psi;t]$ is concentrated on the unit sphere ($\| \psi \| = 1$), and it is constant for all the states within the same projective ray, i.e., states that differ only by a phase factor. The probability distribution shows how the density operator $\rho (t)$ is constructed as a statistical mixture of pure states $\rho_\psi = |\psi \rangle \langle \psi |$. With the above representation of the density operator, the average value of any physical observable described by a self-adjoint operator $A$ is given by 
\be
\langle A \rangle = \tr [A \rho (t) ] = \int \rmd \psi P[\psi; t] \langle \psi | A | \psi \rangle = E[ \langle \psi | A | \psi \rangle ], 
\ee
which is the ensemble average of a quantum mechanical average $\langle \psi | A | \psi \rangle$.

\begin{figure}[t]
\begin{center}
\includegraphics{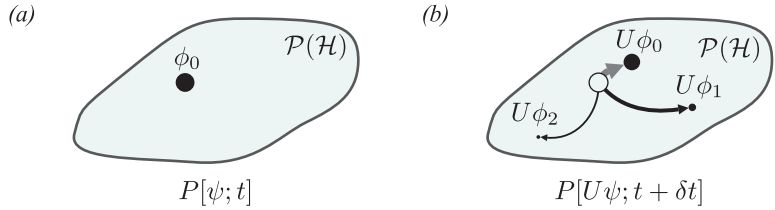}
\caption{\label{fig:evolution} Two-fold evolution in the projective Hilbert space $\mathcal{P} ( \mathcal{H} )$. \textsl{(a)} At time $t$ the probability density $P[\psi;t]$ is concentrated on state $| \phi_0 \rangle$. \textsl{(b)} Evolution over a small time step $\delta t$ propagates the initial state $| \phi_0 \rangle$ (open circle) continuously in $\mathcal{H}$ to state $U | \phi_0 \rangle$ (along the thick grey arrow). On the other hand, the probability density is modified to $P[U\psi;t +\delta t]$, such that the initial weight of state $|\phi_0\rangle$ is redistributed between the determistically evolved $U | \phi_0 \rangle$ and other states $U | \phi_1 \rangle$ and $U | \phi_2 \rangle$ (thin black arrows).}
\end{center}
\end{figure}

Describing the density operator in terms of pure states allows us to interpret the evolution as a two-fold process. Let us consider an infinitesimal time interval $[t,t+\delta t)$. The density operator evolves during this period as $\rho (t) \mapsto \rho( t+\delta t) = \rho(t) + \delta\rho (t)$, where the increment 
\be
\delta \rho (t) = \delta t \frac{ \rmd \rho(t)}{ \rmd t }
\label{eq:masterEquationCondition}
\ee
is given by the master equation~\eref{eq:masterEquation} in the limit of $\delta t \to 0$. On the other hand, the pure-state decomposition~\eref{eq:density} of the density operator, being constructed of two parts, gives
\be
\delta \rho (t) = \int d \psi \big( \delta P[\psi; t] \rho_\psi + P[\psi;t] \delta \rho_\psi (t) \big).
\label{eq:delta_rho}
\ee
This allows an interpretation of the evolution as (i) drift of the pure states $\rho_\psi \mapsto \rho_{U\psi} = \rho_\psi + \delta \rho_\psi (t)$ and (ii) drift of probability $P[\psi;t] \mapsto P[U\psi;t+\delta t] = P[\psi;t] + \delta P[\psi;t]$ among the pure states, such that $\rho(t+\delta t) = \int \rmd\psi P[U\psi; t+\delta t] \rho_{U\psi}$. Figure~\ref{fig:evolution} illustrates the two processes involved. The task of this paper is to derive the pair of $\delta P[\psi;t]$ and $\delta \rho_\psi (t)$ such that the identity~\eref{eq:masterEquationCondition} holds with the master equation~\eref{eq:masterEquation}. From now on, the time arguments are omitted for simplicity. 

\subsection{Drift of pure states}

Expressing the pure-state increment as $\delta \rho_\psi = |\delta \psi \rangle \langle \psi | + |\psi \rangle \langle \delta \psi |$, the structure of the master equation~\eref{eq:masterEquation} suggests a specific form for the evolution $|\psi \rangle \mapsto U(t+\delta t, t)|\psi\rangle = |\psi\rangle + |\delta \psi  \rangle$: gathering terms acting symmetrically from left and right, the evolution is given by $U'(t+\delta t, t) = 1 -\rmi \delta t H_\textrm{eff} /\hbar$, where the non-Hermitian Hamiltonian
\be
H_\textrm{eff}  = H  - \frac{\rmi \hbar}{2} \sum_k \Delta_k  C_k^\dagger  C_k 
\label{eq:Hamiltonian}
\ee
is familiar from the MCWF methods \cite{Dalibard1992,Dum1992,Carmichael1993}. The operation of $U'$ is non-unitary and does not preserve the normalization. However, this can be compensated by defining a propagator $U$ as $U(t+\delta t,t) |\psi\rangle = U'(t+\delta t, t) |\psi\rangle / \| U'(t+\delta t, t) \psi \|$, which gives in the first order of $\delta t$
\be
| \delta \psi  \rangle = \delta t \Big( - \frac{ \rmi }{ \hbar } H_\textrm{eff}  + \frac{1}{2} \sum_k \Delta_k  \| C_k  \psi \|^2 \Big) |\psi \rangle.
\label{eq:deltaPsi}
\ee
As a result, adding a non-linear term to $H_\textrm{eff}$ confines the state $| \psi \rangle$ to the unit sphere of the Hilbert space \cite{Breuer1995a}. Consequently, the pure-state density operator $\rho_\psi$ evolves during an infinetesimal time increment $\delta t$ by
\be
\delta \rho_\psi =  \delta t \Big(  - \frac{ \rmi }{ \hbar } \big[ H , \rho_\psi \big] - \frac{1}{2} \sum_{k}  \Delta_k  \big\{ C_k^\dagger  C_k , \rho_\psi \big\} + \rho_\psi \sum_{k} \Delta_k  \| C_k  \psi \|^2 \Big). 
\label{eq:delta_rho_psi}
\ee
It is straightforward to verify that $\tr [\delta \rho_\psi ] = 0$ and $\delta \rho_\psi  \to 0$ as $\delta t \to 0$, as requested.

\subsection{Drift of probability}

The probability density drift $\delta P$ can now be solved from equation~\eref{eq:delta_rho} by demanding the identity~\eref{eq:masterEquationCondition} with the master equation~\eref{eq:masterEquation}, and by substituting the state increment \eref{eq:delta_rho_psi} therein. Within the pure-state decomposition \eref{eq:density}, the condition becomes
\begin{eqnarray}
&& \int \rmd\psi \Big( \delta P[ \psi] + \delta t \sum_{k} P[ \psi ] \Delta_k  \| C_k  \psi \|^2 \Big) \rho_\psi \nonumber \\
& = & \delta t \sum_{k} \Delta_k  C_k  \rho  C_k^\dagger = \delta t \int \rmd\phi \sum_{k} P[ \phi ] \Delta_k  C_k  \rho_{\phi} C_k^\dagger .
\label{eq:condition1}
\end{eqnarray}
Consequently, using a delta-functional on Hilbert space, defined by $\int \rmd\psi \delta [| \psi \rangle - | \phi \rangle] F[ \psi] = F[\phi ]$, where $F$ is an arbitrary smooth functional, the condition~\eref{eq:condition1} can be written in form
\begin{eqnarray}
\fl \int \rmd\psi \Big( \delta P[\psi] + \delta t \sum_{k} P[ \psi ] \Delta_k  \| C_k  \psi \|^2 \nonumber \\
- \delta t \int \rmd\phi \sum_{k} P[ \phi ] \Delta_k  \| C_k  \phi\|^2 \delta \Big[ | \psi \rangle - \frac{C_k  |\phi \rangle }{\| C_k  \phi \|} \Big] \Big) \rho_{\psi} \nonumber \\
\equiv \int \rmd\psi Q[\psi,t,\delta t] \rho_\psi = 0.
\label{eq:condition2}
\end{eqnarray}
Here, as well as in the rest of the paper, the phase difference between the states appearing in the delta-functional is arbitrary and discarded. Since \eref{eq:condition2} is in the form of the pure-state decomposition, the distribution functional $Q$ has to vanish identically for all states $| \psi \rangle$ and moments of time $t$. Therefore, the probability drift during the time interval $[t,t+\delta t)$ is given by
\begin{eqnarray}
\fl \delta P[\psi] = \delta t \int \rmd\phi \sum_{k} P[ \phi ] \, \Delta_k  \| C_k  \phi \|^2 \delta \Big[ | \psi \rangle - \frac{C_k  |\phi \rangle }{\| C_k  \phi \|} \Big] - \delta t \sum_{k} P[ \psi ] \Delta_k  \| C_k  \psi \|^2 .
\label{eq:deltaP}
\end{eqnarray}

The overall probability is conserved, since $\int \rmd\psi \delta P [\psi] = 0$, so any increase of the density in one region of the state space is always compensated by a decrease elsewhere; the connection between the regions is given by the delta-functional. The phase-invariance of $P$ is conserved, and the limit $\delta P[\psi ] \to 0$ as $\delta t \to 0$ guarantees the continuity in time.

\section{\label{sec:connection}Analysis of the jump processes and their connection to NMQJ method}

The probability drift $\delta P$ connects separate parts of the Hilbert space and, therefore, corresponds to a jump-like evolution. These jumps connect a \textsl{source state} to a \textsl{target state}. Therefore, the jump processes can be understood as a mapping $P_S \mapsto P_T$, where $P_S$ and $P_T$ are probability density functionals for the source ($S$) and the target ($T$) states. Allowing all the states to make a quantum jump means that $P_S[ \psi ] = P[\psi]$, where $P$ is the momentary probability density functional describing the density operator in \eref{eq:density}. 

Since the description coincides with the earlier definition, $P[\psi;t] \mapsto P[\psi; t ] + \delta P [\psi; t]$, one can deduce that 
\be
\delta P [\psi] = P_T[ \psi ] - P_S[ \psi ].
\ee
The source and the target state probability densities can be expressed as the marginals of a joint probability density $P_{T,S}$, such that $P_T [\psi ] = P_{T,S} [ \psi, \mathcal{ H } ] = \int \rmd \phi P_{T,S} [\psi, \phi]$ and $P_S [\psi ] = P_{T,S} [ \mathcal{ H }, \psi ]$, which gives 
\be
\delta P [\psi ] = \int \rmd \phi \big( P_{T,S} [ \psi, \phi ] - P_{T,S} [ \phi, \psi ] \big).
\label{eq:Pdifference}
\ee
Furthermore, by Bayes' theorem, $P_{T,S} [ \phi, \psi ] = P_S [ \psi ] P_{T,S}[ \phi | \psi ]$, which defines the conditional probability density functional for jumping to the target state $| \phi \rangle$ given that the source state is $| \psi \rangle$. 

Writing the probability increment \eref{eq:deltaP} as
\begin{eqnarray}
\fl \delta P[\psi] = \int \rmd\phi \, \delta t \sum_{k} \Big( P[ \phi ] \, \Delta_k \| C_k \phi \|^2 \delta \Big[ | \psi \rangle - \frac{C_k |\phi \rangle }{\| C_k \phi \|} \Big] \nonumber \\
- P[ \psi ] \Delta_k \| C_k \psi \|^2 \delta \Big[ | \phi \rangle - \frac{C_k |\psi \rangle }{\| C_k \psi \|} \Big] \Big)
\label{eq:deltaP_antisymmetric}
\end{eqnarray}
shows that the total probability drift is indeed given by an integral over a functional which is antisymmetric with respect to swapping the arguments. Evidently, it is not at all unambiguous to form the joint probability density $P_{T,S}$ based on the knowledge about the difference of its marginals. Adding an arbitrary antisymmetric functional $F[\psi,\phi] = f[\psi,\phi] - f[\phi,\psi]$, for which $\int \rmd\phi F[\psi,\phi] = 0$, to the integrand of \eref{eq:deltaP_antisymmetric} does not affect the difference $\delta P$ in any way even though it would correspond to some jump processes. This extra term could describe, e.g., cyclic drift $| \psi \rangle \to | \phi \rangle \to | \phi' \rangle \to | \psi \rangle$, which would not have any net effect on any state even though pairwise net drift $| \psi \rangle \leftrightarrow | \phi \rangle $ and, hence, $P_{T,S} [\phi,\psi]$ would be different. In any case, some non-zero $f$ has to be added in order to make the connection between the properly normalized probability density $P_{T,S}$ and the terms appearing in the integrand of \eref{eq:deltaP_antisymmetric}. 

The form of \eref{eq:deltaP_antisymmetric} shows that each decay channel $k$ contributes to the total drift independently. On the other hand, each term corresponding to a single channel $k$ is on its own antisymmetric. The simplest possible way to proceed is to gather up the positive contributions by each channel together and to select $f[\phi,\psi]$ such that it corresponds to a complementing trivial process $| \psi \rangle \mapsto | \psi \rangle$. It will turn out that this choice gives the connection to the probabilities used in the NMQJ (and MCWF) method.

With the above selection, the joint probability density can be written as
\be
P_{T,S} [ \phi, \psi ] = \sum_k P_{T,S}^k [\phi, \psi ] + \tilde{P}_{T,S} [\phi, \psi ],
\label{eq:P_joint}
\ee
where $P_{T,S}^k$ is the positive contribution of each channel $k$ and $\tilde{P}_{T,S} [\phi,\psi] \propto \delta [ | \phi \rangle - | \psi \rangle ]$ is the trivial process chosen such that the density $P_{T,S}$ is normalized to unity. In the non-Markovian case the decay rates $\Delta_k$ may have periods of negative values. On the other hand, all the other components of the products appearing in the integrand of \eref{eq:deltaP_antisymmetric} (timestep $\delta t$, probability density $P$, norm, and the delta-functional) are by definition non-negative, so the sign of the decay rate alone determines the sign of each product term. Therefore, writing the decay rates in terms of positive and negative parts, $\Delta_k = \Delta_k^+ - \Delta_k^-$, where $\Delta_k^\pm = ( |\Delta_k| \pm \Delta_k )/2 \ge 0$ \cite{Breuer2009}, allows gathering up the positive and the negative components from \eref{eq:deltaP_antisymmetric}. This gives
\begin{eqnarray}
\fl P_{T,S}^k [\phi, \psi ] = \delta t \Delta_k^+ P[\psi ] \| C_k \psi \|^2 \delta \Big[ | \phi \rangle - \frac{ C_k | \psi \rangle }{\| C_k \psi \|} \Big] \nonumber \\
+ \delta t \Delta_k^- P[\phi ] \| C_k \phi \|^2 \delta \Big[ | \psi \rangle - \frac{ C_k | \phi \rangle }{\| C_k \phi \|} \Big].
\label{eq:Pk_joint}
\end{eqnarray}
Applying Bayes' theorem defines corresponding positive conditional jump probability components for each channel $k$ by identity $P_{T,S} [\phi,\psi] = P_S[\psi] P_{T,S} [\phi|\psi] = \sum_k P_S[\psi] P_{T,S}^k [\phi|\psi] + P_S[\psi] \tilde{ P }_{T,S} [\phi|\psi]$. Therefore, using the earlier identity $P_S = P$, the result is
\begin{eqnarray}
\fl P_{T,S}^k [\phi| \psi ] = \delta t \Delta_k^+ \| C_k \psi \|^2 \delta \Big[ | \phi \rangle - \frac{ C_k | \psi \rangle }{\| C_k \psi \|} \Big] \nonumber \\
+ \delta t \Delta_k^- \frac{ P[\phi ] }{ P[ \psi ] } \| C_k \phi \|^2 \delta \Big[ | \psi \rangle - \frac{ C_k | \phi \rangle }{\| C_k \phi \|} \Big],
\label{eq:Pk_conditional}
\end{eqnarray}
and the additional trivial operation, given by the normalization rule $P_{T,S} [\mathcal { H } | \psi] = \sum_{k } P_{T,S}^k [ \mathcal{ H } | \psi ] + \tilde{P}_{T,S} [ \mathcal{ H } | \psi ]= 1 $, is then
\be
\tilde{P}_{T,S} [\phi | \psi ] = \bigg( 1 - \sum_{k} \int \rmd \phi' P_{T,S}^k [ \phi' | \psi ] \bigg) \delta [ |\phi \rangle - | \psi \rangle ].
\ee
With this definition, also the joint probability density is automatically normalized to unity: $P_{T,S} [ \mathcal{H}, \mathcal{H} ] = 1$. 

The conditional jump probabilities~\eref{eq:Pk_conditional} are for (time-dependent) Markovian case ($\Delta_k^+ \ge 0$ and $\Delta_k^- = 0$ always) exactly those of the MCWF method \cite{Breuer1995a,Breuer1995b}. Non-Markovian systems allow $\Delta_k^- > 0$ (during which $\Delta_k^+ = 0$), and then the given conditional jump probability coincides with the ones used in the NMQJ method \cite{Piilo2008,Piilo2009}. The trivial operation probability $\tilde{ P }_{T,S}$ vanishes in the integration of equation~\eref{eq:Pdifference} and, therefore, corresponds to not making any quantum jumps but, in the first order of $\delta t$, deterministic evolution instead.

\section{\label{sec:discussion}Discussion}

Looking at a pair of states, $| \psi \rangle$ and $| \phi \rangle$, the decay channels allow jump-like transitions between the two with probabilities derived in the previous section. Moreover, looking at the dynamics generated by a single decay channel $k$, there is a pairwise symmetry in the derived probabilities~\eref{eq:Pk_joint}:
\be
P_{T,S}^k [\phi, \psi ] \Big|_{\Delta_k = -\Delta} = P_{T,S}^k [\psi, \phi ] \Big|_{\Delta_k = + \Delta}.
\ee
This means that every process $|\psi \rangle \to |\phi \rangle$ induced by channel $k$ is replaced by an inverse process $|\phi \rangle \to |\psi \rangle$ occurring with the exactly same \textsl{joint} probability, in case the sign of the decay rate is reversed. 

In an ensemble unravelling there is a set of $N$ states, $\{ |\psi_i \rangle \}$, which evolve stochastically according to the probabilities given in the previous section. The ensemble estimates the probability density functional by $P[\psi ] \simeq (1/N) \sum_i \delta[ | \psi \rangle - | \psi_i \rangle ] = (1/N) \sum_j N_j \delta [ | \psi \rangle - | \phi_j \rangle ]$, where $N_j = \#\{| \psi_i \rangle | | \psi_i \rangle \sim | \phi_j \rangle \}$ counts the number of ensemble members within the same projective ray as $|\phi_j\rangle$, such that $\sum_j N_j = N$. In~\cite{Piilo2009} the set $\{ | \phi_j \rangle \}$ is called the effective ensemble. Each state $|\psi_i \rangle$, and equivalently $|\phi_j \rangle$, is provided with a set of conditional jump probabilities $\{ P_{T,S}^k [\phi | \psi_i ] \} \cup \{ \tilde{ P }_{T,S} [\phi | \psi_i ] \}$ corresponding to the mutually exclusive choice between using one of the decay channels $k$ or the complementing trivial no-jumps option. With the given probabilities the state transforms to state $|\phi \rangle$. 

With $\Delta_k > 0$, the connection between the states given by the delta-functional is $|\phi \rangle \sim C_k | \psi_i \rangle / \| C_k \psi_i \|$ (the relative phase is arbitrary). This correspondence is of one-to-one type in the projective space since the condition defines the ray of $|\phi \rangle$ unambiguously for a given $| \psi_i \rangle$. In the MCWF method the jump between the rays is performed by a non-linear operation $|\psi_i \rangle \mapsto C_k | \psi_i \rangle / \| C_k \psi_i \| \sim | \phi \rangle$. Equivalently, the same result is achieved by opearating with an ensemble-dependent linear jump operator $A_k [\psi_i] \equiv C_k | \psi_i \rangle \langle \psi_i | / \| C_k \psi_i \|$ on the source state $|\psi_i\rangle$. 

In the non-Markovian regime with $\Delta_k < 0$, the conditional jump probability becomes ensemble dependent through the ratio $P[\phi] / P[\psi_i]$, and the connection between the states is reversed: $|\psi_i \rangle \sim C_k | \phi \rangle / \| C_k \phi \|$. Compared with the positive-decay-rate case above, the ray of the target state is now ambiguous, since the condition can be fulfilled by states $|\phi\rangle$ which are not mutually equivalent and which, therefore, correspond to different values of $P[\phi]$. Consequently, there can be many target states $|\phi \rangle$ for which the conditional jump probability for a given \textsl{single} decay channel $k$ is non-zero. Moreover, the mapping between the states $|\psi_i \rangle$ and $| \phi \rangle$ can not be given by the plain Lindblad operator $C_k$ now, but rather by the adjoint of the earlier defined ensemble-dependent jump operator $A_k ^\dagger[\phi] = | \phi \rangle \langle \phi | C_k^\dagger / \| C_k \phi \| $ acting on the source state $|\psi_i\rangle$. Since the jump probability is proportional to $P[\phi]$, the operators $A_k [\phi]$ indeed correspond to states which are present in the ensemble. 

\begin{figure}[t]
\begin{center}
\includegraphics{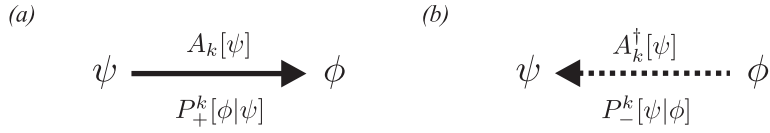}
\caption{\label{fig:conjugateProcesses} \textsl{(a)} The joint probability to jump from a source state $| \psi \rangle$ to a target state $| \phi \rangle$ using a channel $k$, with $\Delta_k = \Delta > 0$, is given by $P[\psi] P_+^k [\phi|\psi]$. Operator $A_k [\psi]$ performs the jump. \textsl{(b)} In the case of an opposite decay rate, $\Delta_k= -\Delta$, a dual process in the opposite direction is applied causing an equally strong but opposite probability stream $P[\phi] P_-^k[\psi|\phi]$.}
\end{center}
\end{figure}

In conclusion, a process defined by the operator $A_k[\psi]$ and probability $P_+^k [ \phi | \psi ] = P_{T,S}^k [\phi | \psi ]\big|_{\Delta_k > 0}$ gets replaced by its dual counterpart characterized by the adjoint operator $A_k^\dagger [\psi]$ and probability $P_-^k [ \psi | \phi ] = P_{T,S}^k [\psi | \phi ]\big|_{\Delta_k < 0}$, as the decay rate $\Delta_k$ goes negative. This duality is illustrated in figure~\ref{fig:conjugateProcesses}. In the non-Markovian case both the conditional jump probabilities and the corresponding jump operators are ensemble dependent. 

\begin{figure}[t]
\begin{center}
\includegraphics{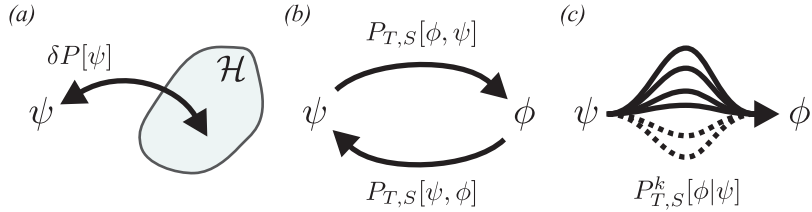}
\caption{\label{fig:probabilityLevels} \textsl{(a)} The total probability drift $\delta P[\psi]$ corresponding to the projective ray of $| \psi \rangle$ during a time interval $[t, t+\delta t)$ is exact for a given deterministic propagator. \textsl{(b)} For a pair of states, $| \psi \rangle$ and $| \phi \rangle$, the net exchange consists of two opposing fluxes of probability $P_{T,S} [\phi,\psi]$ and $P_{T,S} [\psi,\phi]$. The form of $P_{T,S}$ is ambiguous for a given $\delta P$. \textsl{(c)} The joint probability $P_{T,S}$ divides into mutually exclusive channels $k$ corresponding to the set of Lindblad operators in the master equation. Each channel provides the state with a conditional jump probability and a jump operator. Channels with positive (solid arrows) and negative (dashed arrows) decay rates have different but related probability-operator pairs (cf. figure~\ref{fig:conjugateProcesses}).}
\end{center}
\end{figure}

When looking at the drift of probability density, there has been essentially three different levels of detail under consideration. The overall probability drift $\delta P[\psi]$ corresponding to the ray of $| \psi \rangle$ is given by \eref{eq:deltaP}, and it is for a chosen deterministic propagator $U$ exact. This level is illustrated in figure~\ref{fig:probabilityLevels}\textsl{(a)}. However, the possibility of using some different kind of a propagator and thereby accompanying different form of a probability drift is still left open. At the second level of detail the total probability drift $\delta P [\psi ]$ is expressed as a difference of pairwise processes between states $| \psi \rangle$ and $| \phi \rangle$, with fluxes given by the joint probabilities $P_{T,S} [\phi,\psi]$ and $P_{T,S} [\psi,\phi]$ [see equation~\eref{eq:Pdifference} and figure~\ref{fig:probabilityLevels}\textsl{(b)}]. This division was shown to be ambiguous and to allow arbitrary processes with zero net effect but non-zero pairwise net flux. Neglecting the possible non-trivial processes with vanishing net effect and dividing the overall probability flux $P_{T,S} [\phi,\psi]$ from the source state $| \psi \rangle$ to the target state $| \phi \rangle$ into separate terms, which correspond to the set of decay channels in the master equation~\eref{eq:masterEquation}, is the third level of detail [see equation~\eref{eq:P_joint} and figure~\ref{fig:probabilityLevels}\textsl{(c)}]. Especially, breaking the joint probability concerning the whole ensemble by Bayes' theorem to conditional jump probabilities provides each ensemble member with a set of jump probabilities. This is the level where the MCWF and NMQJ methods operate. 

As a last remark, let us discuss the non-trivial question concerning positivity in the non-Markovian dynamics. According to the proof given in \cite{Breuer2009}, the violation of positivity, which manifests the breakdown of the approximation scheme used during the derivation of the master equation, necessarily reflects to a singularity in the conditional reverse-jump probability. In the terminology of this paper, at the moment the density operator becomes negative, there are necessarily a channel $k_0$ and states $| \psi \rangle$ and $| \phi \rangle$, such that $\Delta_{k_0}<0$ and $P_{T,S}^{k_0} [ \phi, \psi ] > 0$ even though $P_S [\psi] = 0$. This is in conflict with Bayes' theorem, and the algorithm terminates. However, the breakdown of the NMQJ method at the negativity is only an implication and the reverse does not need to be true. 

The key element is the assumption that the algorithm always produces a proper probability density functional $P$ within the set of basis states generated by the algorithm itself (the support of the probability density functional, $\{ |\psi \rangle \in \mathcal{H} | P[ \psi ] \neq 0 \}$, is understood as the basis set since the density operator is expressed as a statistical mixture of such states). Given a pure initial state $\rho (0) = |\psi_0 \rangle \langle \psi_0 |$, corresponding to a point spectrum $P[\psi;0] = \delta[ |\psi \rangle - | \psi_0 \rangle ]$, the basis set at later times consists of all the states reachable via different combinations of operations by propagator $U$ and jump operators $C_k$ at any consecutive times on state $|\psi_0 \rangle$: $U(t,t_n ) C_{k_n} (t_n) U(t_n,t_{n-1} ) \cdots C_{k_1} (t_1) U(t_1,0) |\psi_0 \rangle$. Naturally, these states do not generally correspond to the momentary eigenvectors of the density operator, and hence the weights of such states are not necessarily non-negative even though $\rho$ would be positive. Therefore, the question arises, whether it could be possible to generate the dynamics in terms of quasiprobability densities, for which $P$ may have also negative values. Clearly, equations \eref{eq:delta_rho_psi} and \eref{eq:deltaP} give the increments that could be used for direct integration within the projective Hilbert space without the estimation by a stochastic process. If the dynamics is confined to a restricted number of different pure states (cf.~examples in \cite{Piilo2009}) this method would be also efficient, while in the general case a stochastic process would be needed to restrict the number of different states.

\section{\label{sec:summary}Summary and conclusions}

The dynamics of an open quantum system described by a general time-local non-Markovian master equation was analyzed. Using a pure-state decomposition the dynamics was described as (i) deterministic evolution and (ii) evolution of the corresponding probability density functional in the projective Hilbert space. This division allows the interpretation of the dynamics as a continuous drift in Hilbert space accompanied with jump-like transitions between different parts of the state space. 

Certain assumptions and choices were made in order to recover the NMQJ method: (i) the deterministic evolution is given by a certain specific propagator, (ii) probability density functional describing the density operator maps to a valid probability density functional during every time step, (iii) each decay channel $k$ contributes independently and the choice between the channels is mutually exclusive, and (iv) the remaining jump probability is complemented by a trivial identity operation corresponding to a no-jump option. The first choice is motivated by the symmetry issues and it coincides with the one used in the MCWF method, the extension of which is the NMQJ method. For the used deterministic propagator, the corresponding probability drift is exact. Using the second assumption, the probability drift is given by a difference between two marginals of a joint probability distribution for source and target states. The third and the fourth assumptions are motivated purely by simplicity.

The reverse jumps \cite{Piilo2008,Piilo2009} were formulated in terms of ensemble-dependent jump operators. In contrast to the Markovian case, both the conditional jump probabilities and the corresponding jump operators are in the non-Markovian case ensemble dependent, which can be seen as a manifestation of the memory effects. 

Open questions remain for future investigations. The NMQJ method terminates as soon as the conditional jump probability diverges irrespective of whether the density operator is actually about to lose its positivity or not. Could some of the listed assumptions perhaps be relaxed in order to allow a more general stochastic simulation? Or more generally, is there any stochastic process within the reduced system's Hilbert space that is able to simulate the general time-local master equation whenever it is physically valid?

\ack

I thank J. Piilo for interesting discussions and K.-A. Suominen for useful comments. This work was supported by the National Graduate School of Modern Optics and Photonics and the Magnus Ehrnrooth Foundation. Also, support through the Academy of Finland (Projects No. 115682 and 115982) is appreciated.


\section*{References}
\begin{thebibliography}{10}
\bibitem{Breuer2007} Breuer~H-P and Petruccione~F 2007 \textsl{The theory of open quantum systems} (Oxford: Oxford University Press)
\bibitem{Strunz1999} Strunz W T, Di\'{o}si L, and Gisin N 1999 \textsl{Phys. Rev. Lett.} \textbf{82} 1801
\bibitem{Gardiner1999} Gardiner C W and Zoller P 1999 \textsl{Quantum Noise} (Berlin: Springer-Verlag)
\bibitem{Lai2006} Lai C W, Maletinsky P, Badolato A, and Imamoglu A 2006 \textsl{Phys. Rev. Lett.} \textbf{96} 167403
\bibitem{Shao2004} Shao J 2004 \textsl{J. Chem. Phys.} \textbf{120} 5053 
\bibitem{Aharonov2006} Aharonov D, Kitaev A, and Preskill J, 2006 \textsl{Phys. Rev.
Lett.} \textbf{96} 050504
\bibitem{Thorwart2009} Thorwart M, Eckel J, Reina J H, Nalbach P, and Weiss S 2009 \textsl{Chem. Phys. Lett.} \textbf{478} 234
\bibitem{Rebentrost2009} Rebentrost P, Chakraborty R, and Aspuru-Guzik A 2009 \textsl{J. Chem. Phys.} \textbf{131} 184102
\bibitem{Maniscalco2006} Maniscalco S, Piilo J, and Suominen K-A 2006 \textsl{Phys. Rev. Lett.} \textbf{97} 130402
\bibitem{Nakajima1958} Nakajima S 1958 \textsl{Progr. Theor. Phys.} \textbf{20} 948 
\bibitem{Zwanzig1960} Zwanzig R 1960 \textsl{J. Chem. Phys.} \textbf{33} 1338
\bibitem{Shibata1977} Shibata F, Takahashi Y, and Hashitsume N 1977 \textsl{J. Stat. Phys.} \textbf{17} 171
\bibitem{Chaturvedi1979} Chaturvedi S and Shibata F 1979 \textsl{Z. Phys.} B \textbf{35} 297 
\bibitem{Piilo2008} Piilo~J, Maniscalco~S, H\"{a}rk\"{o}nen~K, and Suominen~K-A 2008 \textsl{Phys.~Rev.~Lett.}~\textbf{100} 180402 
\bibitem{Piilo2009} Piilo~J, H\"{a}rk\"{o}nen~K, Maniscalco~S, and Suominen~K-A 2009 \textsl{Phys.~Rev.} A~\textbf{79} 062112 
\bibitem{Dalibard1992} Dalibard J, Castin Y, and M{\o}lmer K 1992 \textsl{Phys. Rev. Lett.} \textbf{68} 580
\bibitem{Dum1992} Dum R, Zoller P, and Ritsch H 1992 \textsl{Phys. Rev.} A \textbf{45} 4879
\bibitem{Carmichael1993} Carmichael H 1993 \textsl{An Open Systems Approach to Quantum Optics} (Berlin: Springer-Verlag)
\bibitem{Garraway1997} Garraway B M 1997 \textsl{Phys. Rev.} A \textbf{55} 2290
\bibitem{Breuer1999} Breuer H-P, Kappler B, and Petruccione F 1999 \textsl{Phys. Rev.} A \textbf{59} 1633
\bibitem{Breuer2004} Breuer H-P 2004 \textsl{Phys. Rev.} A \textbf{70} 012106
\bibitem{Breuer2009} Breuer H-P and Piilo J 2009 \textsl{Europhys. Lett.} \textbf{85} 50004
\bibitem{GoriniA} Gorini V, Kossakowski A, and Sudarshan E C G 1976 \textsl{J. Math. Phys. (N.Y.)} \textbf{17} 821
\bibitem{LindbladB} Lindblad G 1976 \textsl{Commun. Math. Phys.} \textbf{48} 119
\bibitem{Breuer1995a} Breuer H-P and Petruccione F 1995 \textsl{Phys. Rev.} E \textbf{51} 4041
\bibitem{Breuer1995b} Breuer H-P and Petruccione F 1995 \textsl{Phys. Rev. Lett.} \textbf{74} 3788
\end{thebibliography}
\end{document}